\documentstyle[12pt]{article}
\begin{document}
\begin{flushright}
hep-th/0409285\\ BHU-SNB/Preprint
\end{flushright}
\vskip 2.5cm
\begin{center}
{\bf \Large {NONCOMMUTATIVITY IN A SIMPLE TOY MODEL }}

\vskip 3.5cm

{\bf R. P. MALIK} \\
 {\it S. N. Bose National Centre for Basic Sciences,}
\\ {\it Block-JD, Sector-III, Salt Lake, Calcutta-700 098, India}\\

\vspace{0.2cm}

and\\

\vspace{0.2cm}

{\it Centre of Advanced Studies, Physics Department,}\\ {\it
Banaras Hindu University, Varanasi-221 005, India}\\ {\bf E-mail
address: malik@bhu.ac.in}

\vskip 1.5cm

\end{center}

\noindent {\bf Abstract}: We discuss various symmetry properties
of the reparametrization invariant toy model of a free
non-relativistic particle and show that its commutativity and
noncommutativity (NC) properties are the artifact of the
underlying symmetry transformations. For the case of the symmetry
transformations corresponding to the {\it noncommutative}
geometry, the mass parameter of the toy model turns out to be
noncommutative in nature. By exploiting the
Becchi-Rouet-Stora-Tyutin (BRST) symmetry transformations, we
demonstrate the existence of this NC and show its cohomological
equivalence with its commutative counterpart. A connection between
the usual gauge symmetry transformations corresponding to the {\it
commutative} geometry and the quantum groups, defined on the phase
space, is also established for the present model at the level of
Poisson bracket structure. We show that, for the NC geometry, such
a kind of quantum group connection does not exist.
\\

\baselineskip=16pt

\vskip .7cm



\newpage

\noindent
{\bf 1 Introduction}\\

\noindent During the past few years, there has been an upsurge of
interest in the study of physics, formulated on the general
$D$-dimensional noncommutative spaces (i.e. $[x_i, x_j] \neq 0; i,
j = 1, 2.....D$). Such an interest has been thriving because of
some very exciting modern advances in the realm of string theory,
matrix models, quantum gravity, black hole physics, etc., which
have brought out a few intriguing aspects of the noncommutativity
(NC) in the spacetime structure (see, e.g., [1-4] and references
therein). This kind of NC has also been discussed in the simpler
settings where the redefinitions (and/or the change of variables)
play a very decisive role in the construction of the NC in the
spacetime. To be more precise, these redefinitions allow a mapping
between the commutative spacetime to the noncommutative spacetime.
Both versions of the spacetime, at times, describe the {\it same}
physical system because of the fact that the dynamical equations
of motion remain unchanged. Many examples of quantum mechanical
models, particle mechanics, Landau problem, etc., belong to this
category where the choices of the gauge, redefinitions, restricted
symmetries, etc., lead to the NC in the spacetime structure (see,
e.g., [5-13] and references therein).

Since time immemorial, the basic concepts of symmetries have
played some notable roles in the development of theoretical
physics. In a recent set of papers [10,11,13], the continuous
symmetries have been exploited for some models to demonstrate the
existence of the NC. The purpose of our present paper is to
exploit a set of non-standard  symmetry transformations for the
toy model of a massive non-relativistic (NR) free particle and
show the existence of a NC in the spacetime structure. The
insights gained from this toy model might turn out to be quite
useful in the discussion of the more subtle reparametrization
invariant models of the relativistic particles and strings. The
above NR system is rendered reparametrization invariant by hand
because the ``time'' parameter $t$ of this model is treated on a
par with the space variable $x$ (see, e.g., [14] for details). As
a consequence, both these variables constitute the configuration
space in which a trajectory of the above toy model is parametrized
by a new evolution parameter $\tau$. This model is shown to be
endowed with the first-class constraints in the language of
Dirac's prescription for the classification of constraints [15].
These constraints generate a gauge symmetry transformation which
turns out to be equivalent to the reparametrization symmetry
transformation for the model in a certain specific limit. This
equivalence occurs because of the fact that both the above
symmetry transformations owe their origin to the same
constraint(s) (cf. Sect. 3 below).

A variant of the above gauge symmetry transformation, that leads
to the existence of the NC, has been christened by us as the
non-standard symmetry transformation because it is {\it not}
generated by the first-class constraints of the theory and the
transformations for the space and time variables are put by hand
\footnote{The logical reason behind this kind of choice (that
leads to NC in spacetime) is given in Sec. 5 in terms of the BRST
transformations and their cohomological properties. In fact, a
whole set of transformations are allowed by the cohomological
considerations of the BRST transformations but only a small subset
(cf. (4.5) below) corresponds to the symmetry transformation for
the Lagrangian of the present toy model.} to generate the NC in
the transformed frames. The rest of the transformations for the
dynamical variables are derived by demanding the consistency among
(i) the equations of motion, (ii) the expressions for the
canonical momenta, and (iii) the basic transformations for the
space and time variables. The above consistency requirements
enforce this non-standard symmetry transformation to reduce to a
{\it specific form} (cf. (4.5) below) of the usual gauge
transformation. However, in the process, the mass parameter of the
toy model becomes {\it noncommutative} to retain the NC associated
with the original non-standard symmetry transformations
\footnote{It turns out that the mass parameter becomes
noncommutative only with the space variable $x (\tau)$ but it
remains commutative with the other configuration variable $t
(\tau)$. In fact, to retain the NC of the original non-standard
symmetry transformations, the mass parameter remains commutative
with all the dynamical as well as the auxiliary variables of the
theory but it becomes noncommutative with the space variable.}.
Quite independently, the above specific form of the gauge
transformations (cf. (4.5)) can also be obtained directly from the
gauge symmetry transformations that (i) are generated by the
first-class constraints, and (ii) correspond to the {\it
commutative} geometry. Thus, the NC and commutativity, present in
this model, are different facets of the continuous symmetry
transformations. This observation is consistent with  the result
of [9] where the equivalence between the NC and commutativity for
the present toy model has been demonstrated in the language of the
Dirac bracket formalism for two different choices of the gauge
conditions (which, in turn, are connected with each-other by a
specific kind of gauge transformation). To the best of our
knowledge, the reason behind the existence of such kind of an
equivalence between the NC and commutativity has not yet been
discussed in the modern literature by exploiting the continuous
symmetry transformations {\it alone}.

One of the central results of our present paper is to demonstrate,
the above kind of equivalence, by exploiting the
Becchi-Rouet-Stora-Tyutin (BRST) symmetry transformations and
their cohomological considerations. We show, in particular, that
the NC and commutativity of this model are cohomologically
equivalent (cf. (5.7), (5.8) and (4.5) below). We also consider
the connection of the commutativity and NC of this toy model
within the framework of quantum groups, defined in the phase space
of this model. We demonstrate that the commutativity associated
with the gauge symmetry transformations (cf. (3.3) below) can be
captured in the language of the quantum groups in the phase space.
As it turns out, the above commutativity is associated with the
quantum group $SL_{q,q^{-1}} (2)$ when the phase space variables
are transformed under it. This connection is demonstrated at the
level of the Poisson bracket structures alone. We also show that
the special type of NC associated with the non-standard gauge type
transformations (cf. (4.1) below) cannot be captured by the
quantum groups (that have been considered on the phase space).
This is proven, once again, only at the level of the Poisson
bracket structure. The equivalence between the NC and
commutativity, shown by BRST approach, cannot arise in the realm
of the quantum groups because we do not discuss the symmetry
property of the Lagrangian in the framework of the latter. Rather,
we focus only on the Poisson bracket structure.

Our present investigation is interesting and essential on the
following grounds. First and foremost, our present model provides
a very simple setting to discuss the NC in spacetime structure.
This topic, as is obvious from our earlier discussions, is an
active area of research in theoretical high energy physics.
Second, our model is one of the simplest models to possess the
celebrated reparametrization invariance. Thus, its study would
provide some useful insights that would propel us to our central
endeavour of studying the reparametrization invariant models of
(super)gravity and (super)strings. Third, our present study
provides a meeting-ground for the discussion for the two types of
NCs that are associated with (i) the quantum groups, and (ii) the
Snyder-type spacetime considerations. Fourth, the existence of the
NC in our model owes its origin to the continuous symmetry
transformations which are somewhat {\it new} in their contents and
textures. Finally, the key idea of our present investigation has
already been generalized to more interesting systems (see, e.g.,
[22,23]). Thus, our present investigation is the first step
towards our main goal of developing a general theoretical scheme
to produce NCs in the spacetime structure in a consistent fashion.

The contents of our present paper are organized as follows. In
Sec. 2, we discuss the continuous reparametrization symmetry
transformations for the toy model of a free massive
non-relativistic particle. Section 3 is devoted to the discussion
of the gauge symmetry transformations for this system which are
equivalent to the reparametrization transformations in a  specific
limit. We discuss a non-standard set of continuous symmetry
transformations for the variables of the first-order Lagrangian of
this model in Sec. 4. The (anti-)BRST transformations for this toy
model and comments on the cohomological equivalence of the NC and
commutativity, are considered in Sec. 5. Section 6 deals with the
connection between the above commutative and noncommutative
symmetry transformations and the quantum group symmetry
transformations in the phase space of the above toy model where
emphasis is laid on the Poisson bracket structures. Finally, we
make some concluding remarks in Sec. 7 and point out a few future
directions for further investigations.\\

\noindent
{\bf 2 Reparametrization Symmetry Transformations}\\

\noindent
We begin with the action integral ($S_t$) for the $(1 + 1)$-dimensional
system of a free non-relativistic (NR) particle with mass $m$, as given
below
$$
\begin{array}{lcl}
 S_t = {\displaystyle \int}
dt\; L^{(t)}_{0} (x, \dot x) \equiv
 {\displaystyle \int}\; dt\;
 {\displaystyle \frac{m \dot x^2}{2}},
\end{array} \eqno(2.1)
$$
where the Lagrangian $L^{(t)}_{0} (x,\dot x)$, for the free NR particle,
depends only on the square of the
velocity ($\dot x = \frac{d x}{dt}$) variable $\dot x$ that is  constructed
from the displacement vector $x$ and the evolution parameter ``time'' $t$.
It is evident that the above action is not endowed with the reparametrization
symmetry. However, this symmetry can be brought in by treating
the ``time'' parameter $t$ as a dynamical variable $t(\tau)$ on a par with
the displacement variable $x(\tau)$ (see, e.g., [14]) where $\tau$ is the
new evolution parameter. The action integral $S_t$ (integrated
over the element $dt$ with Lagrangian $L_{0}^{(t)} (x,\dot x)$) of (2.1) can be
transformed to the action integral
$S_\tau$ (integrated over the element $d\tau$ with Lagrangian
$L_{0}^{(\tau)} (x,\dot x,t,\dot t)$) as (see, e.g., [9])
$$
\begin{array}{lcl}
S_\tau = {\displaystyle \int} \; d\tau\; L_{0}^{(\tau)} (x, \dot x, t, \dot t)
\equiv {\displaystyle \int}\; d \tau\;
 {\displaystyle \frac{m \dot x^2}{2\;\dot t}},
\end{array} \eqno(2.2)
$$ where, now, we have $\dot x = dx/d\tau, \dot t = dt/ d\tau$ and
the Jacobian $J$ (in $dt = J d \tau$) is given by $J = dt/d\tau$.
It can be checked that this system is endowed with the first-class
constraint $$
\begin{array}{lcl}
p_x^2 +  2 m p_t \approx 0,
\end{array} \eqno(2.3)
$$
in the language of Dirac's prescription [15] for the classification
of constraints where the conjugate momenta $p_x (\tau)$ and $p_t (\tau)$,
corresponding to the configuration variables $x(\tau)$
and $t(\tau)$,  are defined
in terms of the Lagrangian function $L_{0}^{(\tau)} (x, \dot x, t, \dot t)$, as
$$
\begin{array}{lcl}
p_x = {\displaystyle \frac{\partial L^{(\tau)}_{0}}{\partial \dot x}}
\equiv {\displaystyle \frac{m \dot x}{\dot t}},
\qquad
p_t = {\displaystyle \frac{\partial L^{(\tau)}_{0}}{\partial \dot t}}
\equiv - {\displaystyle \frac{m \dot x^2}{2 \dot t^2}}.
\end{array} \eqno(2.4)
$$
The symbol $\approx$ in (2.3) stands for the weak equality
(see, e.g., [15] for details). The first-order ($L_f^{(\tau)}$)- and the
second-order ($L_s^{(\tau)}$) Lagrangians
can be derived from $L_{0}^{(\tau)}$ by exploiting
(i) the Legendre transformation, and (ii)
 the explicit expressions for the momenta $p_x$ and $p_t$ as given in (2.4).
The ensuing  Lagrangians, thus obtained,  are as follows
$$
\begin{array}{lcl}
&&L^{(\tau)}_{f}
= p_x \dot x + p_t \dot t -
{\displaystyle \frac{1}{2}}\;E\; (p_x^2 + 2 m p_t),
\nonumber\\
&& L^{(\tau)}_{s} = {\displaystyle \frac{ \dot x^2}{ 2 E}}
 + {\displaystyle \frac{m \dot x^2}{2 \dot t}}
\Bigl [ {\displaystyle \frac{ E m} {\dot t}} - 1 \Bigr ],
\end{array} \eqno(2.5)
$$
where $E(\tau)$ is a Lagrange multiplier that takes care of the constraint
(2.3). At this stage, a few comments are in order. First, the massless
limit (i.e. $m \to 0$) for the Lagrangian in
(2.1) and/or (2.2) is {\it not} well defined.
However, the Lagrangians in (2.5) do allow the existence of such a limit.
Second, all the Lagrangians $L_0^{(\tau)}, L_f^{(\tau)}$ and $L_s^{(\tau)}$
are equivalent as  can be seen by exploiting the equations of motion
$\dot x = E p_x, \dot t = E m, \dot p_x = 0, \dot p_t = 0$ that emerge
from the first-order Lagrangian $L_f^{(\tau)}$ and the expressions for
the momenta $p_x$ and $p_t$ given in (2.4). It will be noted that the free
motion (i.e. $\dot p_x = 0, \dot  p_t = 0$) of the NR particle leads
to the second-order equations of motion: $\ddot x E - \dot x \dot E = 0,
\ddot x \dot t - \dot x \ddot t = 0$ which emerge from $L_f^{(\tau)}$
and $L_s^{(\tau)}$.
Third, all the three Lagrangians in (2.2) and (2.5) are reparametrization
invariant. For the sake of simplicity
\footnote{In contrast to the Lagrangians $L_0^{(\tau)}$
and $L_s^{(\tau)}$, the first-order Lagrangian $L_f^{(\tau)}$
is simpler in the sense that it has {\it no} variable
(and its derivative(s) w.r.t. $\tau$) in the denominator.
Furthermore, it is endowed with the largest number of variables
(i.e. $x, p_x, t, p_t, E$) that allow more freedom
for theoretical discussions.}, however,
let us check the invariance
of the first-order Lagrangian $L_f^{(\tau)}$ under the infinitesimal
reparametrization transformation $\tau \to \tau^\prime = \tau -
\epsilon (\tau)$ where $\epsilon (\tau)$ is an infinitesimal local parameter.
All the variables
of $L_f^{(\tau)}$ undergo the following change
under the above infinitesimal reparametrization transformation
$$
\begin{array}{lcl}
&&\delta_r x = \epsilon\; \dot x, \qquad
\delta_r t = \epsilon\; \dot t, \qquad
\delta_r p_x  = \epsilon\; \dot p_x, \nonumber\\
&& \delta_r p_t  = \epsilon\; \dot p_t, \qquad
\delta_r E  = {\displaystyle \frac{d}{d\tau}}\; [\epsilon E],
\end{array} \eqno(2.6)
$$
where the transformation $\delta_r$ for a generic variable
$\phi (\tau) \equiv x (\tau), t (\tau), p_x (\tau), p_t (\tau), E (\tau)$
is defined as
$\delta_r \phi (\tau) = \phi^\prime (\tau) - \phi (\tau)$. It is elementary
to verify that the first-order Lagrangian $L_f^{(\tau)}$
remains quasi-invariant under
the transformation $\delta_r$ as it changes to a total derivative
w.r.t. $\tau$ (i.e.
$\delta_{r} L_f^{(\tau)} = (d/d\tau)\; [\epsilon L_f^{(\tau)}]$).
Similarly, one can check that, under (2.6), the other Lagrangians
transform as: $\delta_{r} L_0^{(\tau)} = (d/d\tau)\;
[\epsilon L_0^{(\tau)}]$ and
$\delta_{r} L_s^{(\tau)} = (d/d\tau)\; [\epsilon L_s^{(\tau)}]$.\\

\noindent
{\bf 3  Gauge Symmetry Transformations}\\

\noindent
It is well-known that the existence of the first-class constraints on a
physical system entails upon the Lagrangian of the
system to be endowed with some gauge type of
continuous symmetry transformations. These
symmetry transformations are generated by the first-class constraints
themselves. For instance, for the present toy model described by the
first-order Lagrangian $L_f^{(\tau)}$,
there are only two first-class constraints. These are
explicitly written as $\Pi_{E} \approx 0,
p_x^2 + 2 m p_t \approx 0$ where $\Pi_E$ is the conjugate momentum
corresponding to the Lagrange multiplier variable $E(\tau)$. The
generator $G$ for the gauge transformations can be written,
in terms of the above first-class constraints,  as
$$
\begin{array}{lcl}
G = \Pi_E\; \dot \xi + {\displaystyle \frac{\xi}{2}}\;
(p_x^2 + 2 m p_t),
\end{array} \eqno(3.1)
$$
where $\xi (\tau)$ is the local infinitesimal gauge parameter for
the above transformations. The explicit form of the gauge transformation
$\delta_g$ for the generic local variable $\phi (\tau)
= x (\tau), t (\tau), p_x (\tau), p_t (\tau), E (\tau)$ can be written,
in terms of the above generator $G$, as
$$
\begin{array}{lcl}
\delta_g \; \phi (\tau) = - i\; [\phi (\tau), G ],
\end{array} \eqno(3.2)
$$
where, for the sake of
explicit computation, the canonical commutation relations:
$[ x, p_x ] = i, [t, p_t] = i, [ E, \Pi_E ] = i$ have to be exploited with
the understanding that all the other canonical commutators are zero. Ultimately,
one obtains the following gauge transformations for all the
$\tau$ dependent dynamical variables
$$
\begin{array}{lcl}
\delta_g x = \xi\; p_x, \qquad
\delta_g t = \xi\; m, \qquad
\delta_g p_x  = 0, \qquad
 \delta_g p_t  = 0, \qquad
\delta_g E  = \dot \xi.
\end{array} \eqno(3.3)
$$
Under the above infinitesimal continuous gauge transformations,
the first-order Lagrangian $L_f^{(\tau)}$ remains quasi-invariant because it
changes to a total derivative. This statement can be written explicitly
in the mathematical form as
\footnote{The above gauge transformation
is the symmetry transformation for the other Lagrangians too.
In the explicit form, it can be checked that the
following infinitesimal gauge transformations ($\delta_g^{(0)}$) for the case
of the Lagrangian $L_{0}^{(\tau)}$, lead to:
$\delta_{g}^{(0)} x = \xi\;(\frac{m \dot x}{\dot t}),\;
\delta_{g}^{(0)} t = \xi\;m,\;\delta_{g}^{(0)} L_{0}^{(\tau)} =
\frac{d} {d\tau}[\frac{\xi}{2}\;
 (\frac{m \dot x}{\dot t})^2].$
Similarly, for the second-order Lagrangian $L_s^{(\tau)}$, the infinitesimal
gauge transformations ($\delta_g^{(s)}$) on the dynamical variables
and the Lagrangian yield:
$\delta_{g}^{(s)} x = \xi\;(\frac{\dot x}{E}),\;\delta_{g}^{(s)} t = \xi\;m,\;
\delta_g^{(s)} E  = \dot \xi, \;\delta_{g}^{(s)} L_{s}^{(\tau)} =
\frac{d} {d\tau}\;[\frac{\xi}{2}\; (\frac{\dot x}{E})^2]$.}
$$
\begin{array}{lcl}
\delta_g\; L_f^{(\tau)} = {\displaystyle \frac{d} {d\tau}\;
\Bigl [ \frac{\xi}{2}\; p_x^2 \Bigr ]}.
\end{array} \eqno(3.4)
$$
It will be noted that the continuous gauge transformations (3.3) can also
be obtained from the reparametrization symmetry transformations
(2.6) if we exploit the Euler-Lagrange equations of motion
$\dot x = E p_x, \dot t = E m, \dot p_x = 0, \dot p_t = 0$
(emerging from the first-order Lagrangian
$L_f^{(\tau)}$) and  identify
$ \epsilon (\tau) = \xi (\tau)/ E (\tau)$.  The root cause behind
the existence of the above connection between the continuous gauge- and
reparametrization symmetries
is the fact that {\it both} these symmetries
are generated by the first-class constraints of the theory.

The gauge-transformations (3.3) have been derived by exploiting
the generator $G$, the basic canonical commutators
 and the generic transformation in (3.2).
However, it can be proven non-trivially that these gauge transformations
$\delta_g E = \dot \xi,\; \delta_g p_x = 0, \delta_g p_t = 0$ are correct.
For this purpose, we take, as inputs, only the transformations
for the basic variables $x$ and $t$ and demand their consistency
with the equations of motion and the expressions for momenta in (2.4).
For instance,
the equations of motion $\dot t = E m,\; p_x = \dot x/ E$
and the expression $p_t = - (m \dot x^2)/ (2 \dot t^2)$ imply that
$\delta_g \dot t = m \delta_g E,\; \delta_g p_x = \delta_g (\dot x/ E),
\delta_g p_t = - \frac{m}{2} \delta_g (\dot x^2/\dot t^2)$.
Using $\delta_g t = \xi m,\; \delta_g x = \xi p_x$, we obtain
$\delta_g E = \dot \xi,\; \delta_g p_x = (\dot \xi/ E) (p_x - \dot x/ E)
+ (\xi / E)\; \dot p_x$. However, the equations of motion $\dot p_x = 0$
and $p_x = (\dot x/ E)$, imply that $\delta_g p_x = 0$. Similarly,
the requirement of the consistency among (i) the basic
gauge transformations for $x$ and $t$, (ii) the expressions (2.4), and
(iii) the equations of motion, leads to $\delta_g p_t = 0$. The explicit
computation
$$
\begin{array}{lcl}
 \delta_g p_t  = - {\displaystyle \frac{m \dot x \dot \xi}{\dot t^2}\;
\Bigl (p_x - \frac{m \dot x}{\dot t} \Bigr ) - \frac{m \dot x \xi}{\dot t^2}}
\;\dot p_x,
\end{array} \eqno(3.5)
$$
demonstrates that $\delta_g p_t = 0$ because of the fact that
$\dot p_x = 0$ and $p_x = (m \dot x)/\dot t$. It should be noted that
the transformations $\delta_g p_x = 0, \delta_g p_t = 0$ are also consistent
with the other equations of motion: $\dot p_x = 0, \dot p_t = 0,
p_x^2  + 2 m p_t = 0$. This is due to the fact that $\delta_g \dot p_x = 0,
\delta_g \dot p_t = 0$ and $ 2 p_x \delta_g p_x + 2 m \delta_g p_t = 0$
are automatically satisfied. Thus, we note that the definitions in (2.4),
the gauge transformations in (3.3) and the equations of motion
(deduced from $L_f^{(\tau)}$) are
consistent with one-another.
This non-trivial trick for the derivation of the symmetry transformations
will be exploited in the next section too.\\

\noindent
{\bf 4 Non-Standard Symmetry Transformations and Noncommutativity}\\

\noindent The usual gauge transformations $\delta_g$ in (3.3)
imply that the transformed frame is now characterized by $x^{(g)}
= x + \xi p_x, t^{(g)} = t + \xi m$. With the help of (i) the
canonical Poisson brackets $\{x, p_x\}_{(PB)} = 1, \{t,
p_t\}_{(PB)} = 1, \{x, x\}_{(PB)} = 0, \{ x, t\}_{(PB)} = 0,
\{p_x, p_x\}_{(PB)} = 0, \{t, t\}_{(PB)} = 0, \{p_t, p_t\}_{(PB)}
= 0$, etc., and, (ii) treating the mass parameter to be {\it
commutative} with everything, it can be seen that, in the gauge
transformed  frame too, we have the commutative geometry (i.e.
$[x^{(g)}, t^{(g)}] = 0$) because of the fact that $\{ x^{(g)} ,
t^{(g)} \}_{(PB)} = 0$. To bring in the {\it noncommutative}
geometry, let us focus on some non-standard transformations
$\tilde \delta_g$ $$
\begin{array}{lcl}
x(\tau) \rightarrow x^{(ns)} (\tau) &=&\; x (\tau)+\; \zeta (\tau)
m \;\;\Rightarrow\;\; \tilde \delta_{g} x (\tau) = \zeta (\tau) m,
\nonumber\\ t(\tau) \rightarrow t^{(ns)} (\tau) &=&\; t(\tau)+
\;\zeta (\tau) p_x (\tau) \;\;\Rightarrow \;\; \tilde \delta_{g}
t(\tau) = \zeta (\tau)  p_x(\tau),
\end{array} \eqno(4.1)
$$ where $\zeta (\tau)$ is an infinitesimal transformation
parameter. These transformations have been chosen because (i) they
lead to the NC in the transformed frame due to $\{x^{(ns)},
t^{(ns)} \}_{(PB)} = \zeta (\tau)$, and (ii) they have some
relevance in the context of the nilpotent BRST symmetries and BRST
cohomology (cf. Sec. 5 below). Let us have a close look at (4.1).
Here the infinitesimal increments in the space and time variables
have connections with the corresponding increments in the gauge
transformations (3.3). In fact, the increments of the latter are
exchanged in the former (i.e $\tilde \delta_g x = \delta_g t,
\tilde \delta_g t = \delta_g x$). This is, moreover, consistent
with the gauge choice(s) made in [9]. As we have argued earlier
for the non-trivial way of deriving the usual gauge
transformations (3.3), we derive here, exactly in a similar
fashion, the non-standard transformations $\tilde \delta_g$ for
all the variables of the first-order Lagrangian $L_f^{(\tau)}$.
The basic inputs for such a derivation are the requirements of the
consistency among (i) the equations of motion, (ii) the
expressions for the momenta (2.4), and (iii) the basic
transformations (4.1) for the space and time variables that lead
to the NC. Taking all these considerations into account, we obtain
the following transformations $\tilde \delta_g$: $$
\begin{array}{lcl}
&&\tilde \delta_g  x = \zeta m, \qquad
\tilde \delta_g  t = \zeta p_x, \qquad
\tilde \delta_g  E =
{\displaystyle \frac{1}{m} (\dot \zeta p_x)}, \nonumber\\
&&\tilde \delta_g  p_x = \;\;
- {\displaystyle \frac{ \dot \zeta}{E m}\; \Bigl (\frac{\dot x}{E}\;
p_x - m^2 \Bigr )} \equiv -
{\displaystyle \frac{ \dot \zeta}{E m}\; \Bigl (p_x^2 - m^2
\Bigr )}, \nonumber\\
&&\tilde \delta_g  p_t =
{\displaystyle \frac{m \dot x}{\dot t^2} \; \dot \zeta\; \Bigl (
\frac{\dot x}{\dot t}\; p_x - m \Bigr )} \equiv +
{\displaystyle \frac{p_x}{E m^2} \; \dot \zeta\; \Bigl (p_x^2
- m^2 \Bigr )}.
\end{array} \eqno(4.2)
$$
A few comments are in order now. First, the above transformations have been
deduced by exploiting the basic transformations: $\tilde \delta_g x
= \zeta m, \tilde \delta_g t = \zeta p_x$ and the equations of motion
$\dot x = E p_x,\; \dot t = E m, \dot p_x = 0$ as well as the definition
$p_t = - (m \dot x^2)/ (2 \dot t^2)$. Second, the expressions (2.4) for the
momenta $p_x$ and $p_t$ have been utilized to express the r.h.s. of
the above transformations, ultimately, in terms of the physical
variables $m, E$
and $ p_x$. Third, it can be readily seen that the above transformations
are consistent with the equation of motion $p_x^2 + 2 m p_t = 0$ because
$\tilde \delta_g p_t = - (p_x/m)\; \tilde \delta_g p_x$. Fourth,
the consistency
between the above transformations and the equations of motion $\dot p_x = 0,
\dot p_t = 0$ (i.e. $\tilde \delta_g \dot p_x \equiv (d/d \tau)
\tilde \delta_g p_x = 0,
\tilde \delta_g \dot p_t
\equiv (d/d\tau) \tilde \delta_g p_t = 0$) leads to the following restrictions
$$
\begin{array}{lcl}
&&{\displaystyle \frac{d}{d \tau} \bigl (\tilde \delta_g p_x \bigr )} = 0
\Rightarrow  {\displaystyle \frac{1}{E^2 m} \bigl (p_x^2 - m^2 \bigr )
\bigl (\dot \zeta \dot E - \ddot \zeta E \bigr )} = 0, \nonumber\\
&&{\displaystyle \frac{d}{d \tau} \bigl (\tilde \delta_g p_t \bigr )} = 0
\Rightarrow - {\displaystyle \frac{p_x}{E^2 m^2} \bigl (p_x^2 - m^2 \bigr )
\bigl (\dot \zeta \dot E - \ddot \zeta E \bigr )} = 0.
\end{array} \eqno(4.3)
$$
It is obvious, from the above, that for $E \neq 0, m\neq 0$, we have
the following solutions
$$
\begin{array}{lcl}
(i)\;\; \;p_x\; = \;m,\; \qquad \mbox{and/or}\qquad\;
(ii)\;\;\;\;  \ddot \zeta \;E - \dot \zeta\; \dot E = 0.
\end{array} \eqno(4.4)
$$
Let us focus first on the solution (i) of (4.4). In this
case, the above non-standard transformations $\tilde \delta_g$ reduce to
a new specific transformation $\delta_g^{(sp)}$
(i.e. $\tilde \delta_g \to \delta_g^{(sp)}$), under which, we obtain
the following symmetry
(i.e. $\delta_g^{(sp)} L_f^{(\tau)} = (d/d\tau) [(\zeta m^2)/(2)]$)
transformations
$$
\begin{array}{lcl}
\delta_g^{(sp)} x = \zeta m,\qquad
\delta_g^{(sp)} t = \zeta m,\qquad
\delta_g^{(sp)} E = \dot \zeta,\qquad
\delta_g^{(sp)} p_x = 0,\qquad
\delta_g^{(sp)} p_t = 0.
\end{array} \eqno(4.5)
$$ It is evident that the original NC (i.e. $\{ x^{(ns)}, t^{(ns)}
\}_{(PB)} = \zeta (\tau)$; with $\zeta (\tau) \neq 0$), discussed
earlier under the transformations (4.1), can be retained (for $p_x
= m$) if and only if $\{x, p_x \}_{(PB)} = 1\; \rightarrow \;\{ x,
m \}_{(PB)} = 1$. This demonstrates that the above non-standard
symmetry enforces the mass parameter $m$ of our toy model to
become noncommutative {\it only} with the configuration variable
$x(\tau)$. However, it retains its commutative nature with all the
other variables of the theory. Such kind of NC for the mass
parameter has also appeared for the free motion of a NR particle
on a quantum-line in the framework of ``quantum
groups''[16,17-19]. The solution $p_x = - m$ in (i) of (4.4) does
not correspond to an interesting choice for a free NR ``particle''
because it would mean a negative momentum.

There is a  drastically {\it different} way to look at the
solution $p_x = m$. This is in the language of the gauge
transformations listed in (3.3) which yields the transformations
(4.5) in the limit $p_x = m$ and $\xi (\tau) = \zeta (\tau)$. In
fact, in this limit, one finds the {\it same} transformations for
$x$ as well as $t$ (i.e. $x \to x^{(g)} = x + \zeta m, t \to
t^{(g)} = t + \zeta m$) and, as expected, one obtains the {\it
commutative} geometry (i.e. $\{ x^{(g)}, t^{(g)} \}_{(PB)} = 0$).
It will be noted that, for the standard gauge symmetry
transformations (3.3), the mass parameter $m$ is assumed to be
commutative with everything, right from the beginning. This
establishes the fact that the {\it NC}, present in this toy model,
is only the artifact of the underlying symmetry transformations.
In particular, the symmetry transformations in (4.5) can be
interpreted in {\it two different} ways which lead to NC and
commutativity in the theory. This observation is consistent with
the result obtained in [9]  for this model where the Dirac bracket
considerations have been taken into account for the different
choices of the gauge conditions. In fact, it has been shown
explicitly in [9] that the NC and commutativity for this toy model
owe their origin to the different choices of the gauge conditions
which, in turn, are connected to each-other by a special kind of
gauge transformation.

Finally, let us concentrate on the solution (ii) of (4.4). It is
clear that the solution (ii) implies $(\ddot \zeta/\dot \zeta) =
(\dot E/ E) \equiv R$. This ratio can be chosen in many different
ways. For the choice $ R = \pm K$, we obtain the solutions $\zeta
(\tau) = \zeta (0) e^{\pm K \tau}, E (\tau) = E (0) e^{\pm K
\tau}$ where $K$ is a constant. However, these solutions do not
lead to any interesting symmetry property of the first-order
Lagrangian $L_f^{(\tau)}$. One can choose $R = \pm \tau$, which
leads to $ E(\tau) = E (0) e^{\pm \tau^2/2}$ and $\zeta (\tau) =
\tau \pm \sum_{n = 1}^{\infty} [(2n - 1)!!]/ [(2n + 1)!]\;
\tau^{2n + 1}$. However, these solutions also do not lead to any
interesting symmetry properties of $L_f^{(\tau)}$. At the moment,
we do {\it not} know any solution for $E(\tau)$ and  $\zeta
(\tau)$ that corresponds to any symmetry properties of
$L_f^{(\tau)}$. Thus, we are led to conclude that $p_x = m$ is the
only interesting solution corresponding to a symmetry
transformation of the Lagrangian for our present reparametrization
invariant  toy model and it leads to the existence of the NC in
the spacetime structure.\\

\noindent
{\bf 5 (Anti-)BRST Symmetry Transformations and Noncommutativity}\\

\noindent The ``classical'' local and continuous gauge symmetry
transformations $\delta_g$ in (3.3) can be traded with the
``quantum'' local and continuous gauge transformations which are
nothing but the off-shell nilpotent ($s_{(a)b}^2 = 0$) (anti-)BRST
symmetry transformations $s_{(a)b}$. To obtain the above nilpotent
transformations, the infinitesimal gauge parameter $\xi$ is
replaced by an anticommuting number ($\eta$) and the (anti-)ghost
fields $(\bar c)c$. These off-shell nilpotent ($s_{(a)b}^2 = 0$)
and anticommuting ($s_b s_{ab} + s_{ab} s_b = 0$) (anti-)BRST
transformations $s_{(a)b}$ are \footnote{We follow here the
notations and conventions adopted in [20]. In fact, in its full
blaze of glory, the nilpotent ($\delta_{(a)b}^2 = 0$) (anti-)BRST
transformations $\delta_{(a)b}$ are the product of an
anticommuting spacetime independent parameter $\eta$ (with $\eta c
= - c \eta, \eta \bar c = - \bar c \eta$, etc.) and the
transformations $s_{(a)b}$ (with $s_{(a)b}^2 = 0$).} $$
\begin{array}{lcl}
&&s_b x = c\; p_x, \;\;\qquad s_b t = c \; m, \;\;\qquad s_b p_x =
0, \qquad s_b p_t  = 0, \nonumber\\ && s_b E  = \dot c, \qquad s_b
c = 0, \qquad s_b \bar c = i B, \qquad s_b B = 0,
\end{array} \eqno(5.1)
$$
$$
\begin{array}{lcl}
&&s_{ab} x = \bar c\; p_x, \qquad s_{ab} t = \bar c \; m, \qquad
s_{ab} p_x  = 0, \qquad s_{ab} p_t  = 0, \nonumber\\ && s_{ab} E =
\dot {\bar c}, \qquad s_{ab} \bar c = 0, \qquad s_{ab}  c = -i\;
B, \qquad s_{ab} B = 0,
\end{array} \eqno(5.2)
$$ where $B(\tau)$ is the Nakanishi-Lautrup auxiliary variable.
The above local, continuous and off-shell nilpotent ($s_{(a)b}^2 =
0$) symmetry transformations leave the (anti-)BRST invariant
Lagrangian \footnote{The on-shell ($\ddot c = \ddot {\bar c} = 0$)
nilpotent ($\tilde s_{(a)b}^2 = 0$) (anti-)BRST transformations
$\tilde s_{(a)b}$, namely; $ \tilde s_b x = c\; p_x, \;\tilde s_b
t = c \; m, \;\tilde s_b p_x  = 0, \; \tilde s_b p_t  = 0,\;\tilde
s_b E  = \dot c, \;\tilde s_b c = 0, \; \tilde s_b \bar c = - i
\dot E$ and $\tilde s_{ab} x = \bar c\; p_x, \;\tilde s_{ab} t =
\bar c \; m, \; \tilde s_{ab} p_x  = 0, \; \tilde s_{ab} p_t  = 0,
\; \tilde s_{ab} E  = \dot {\bar c}, \; \tilde s_{ab} \bar c = 0,
\; \tilde s_{ab}  c = + i\; \dot E$ leave the Lagrangian $\tilde
L^{(\tau)}_{b} = p_x \dot x + p_t \dot t - \frac{1}{2}\;E\; (p_x^2
+ 2 m p_t) - \frac{1}{2}\; \dot E^2 - i \dot {\bar c} \dot c$
quasi-invariant. The symmetries $\tilde s_{(a)b}$ and the
Lagrangian $\tilde L_b^{(\tau)}$ are obtained from (5.1), (5.2)
and (5.3) by the substitution $B = - \dot E $ that emerges as the
equation of motion from the Lagrangian (5.3).} (see, e.g., [18]
for the details on such kind of Lagrangians) $$
\begin{array}{lcl}
L^{(\tau)}_{b} = p_x \dot x + p_t \dot t - {\displaystyle
\frac{1}{2}}\;E\; (p_x^2 + 2 m p_t) + B \dot E + {\displaystyle
\frac{1}{2}\; B^2} - i \dot {\bar c} \dot c,
\end{array} \eqno(5.3)
$$
quasi-invariant because this Lagrangian $L_b^{(\tau)}$ transforms as follows
$$
\begin{array}{lcl}
s_b\; L_b^{(\tau)} = {\displaystyle \frac{d} {d\tau}\; \Bigl [
\frac{c}{2}\; (p_x^2 + 2 m p_t) + B \dot c \Bigr ]}, \qquad
s_{ab}\; L_b^{(\tau)} = {\displaystyle \frac{d} {d\tau}\; \Bigl [
\frac{\bar c}{2}\; (p_x^2 + 2 m p_t) + B \dot {\bar c} \Bigr ]},
\end{array} \eqno(5.4)
$$ under the (anti-)BRST transformations $s_{(a)b}$. The
generators for the above off-shell nilpotent (anti-)BRST
transformations are the off-shell nilpotent ($Q_{(a)b}^2 = 0$),
anticommuting ($Q_b Q_{ab} + Q_{ab} Q_b = 0$) and conserved ($\dot
Q_{(a)b} = 0$) (anti-)BRST charges $Q_{(a)b}$ as given below $$
\begin{array}{lcl}
Q_b = B\; \dot c + {\displaystyle \frac{c}{2}} (p_x^2 + 2 m p_t)
\equiv B \dot c - \dot B c, \quad Q_{ab} = B\; \dot {\bar c} +
{\displaystyle \frac{\bar c}{2}} (p_x^2 + 2 m p_t) \equiv B \dot
{\bar c} - \dot B \bar c,
\end{array} \eqno(5.5)
$$ where the Euler-Lagrange equation $\dot B = - \frac{1}{2}
(p_x^2 + 2 m p_t)$, emerging from the (anti-)BRST invariant
Lagrangian (5.3),  has been exploited. A few comments are in order
at this juncture. First, the above expressions in (5.5) are the
generalizations of the expression in (3.1). Second, the nilpotency
of the charges in (5.5) can be proven by exploiting the canonical
(anti-)commutators $ [x, p_x] = i, [t, p_t ] = i, [E, B] = i, \{
c, \dot {\bar c} \} = + 1, \{\bar c, \dot c \} = - 1$ in the
computations $Q_b^2 = \frac{1}{2} \{ Q_b, Q_b \} = 0, Q_{ab}^2 =
\frac{1}{2} \{ Q_{ab}, Q_{ab} \} = 0$. Third, the generic
transformations in (3.2) can be generalized to $s_{(a)b} \phi = -
i [\phi, Q_{(a)b}]_{\pm}$ where the subscripts $(+)-$ on the
square brackets stand for the (anti)commutators for a given
generic variable $\phi$ being (fermionic)bosonic in nature.
Fourth, the physicality criteria ($Q_{(a)b} |phys> = 0$) imply
that the physical states (i.e. $|phys>$) are the subset of the
total Hilbert space of states which are annihilated by the
operator form of the first-class constraints of the theory. This
statement can be succinctly expressed in the mathematical form due
to the requirement that the condition $Q_{(a)b}\; |phys> = 0$
implies the following $$
\begin{array}{lcl}
(i)\; \Pi_E (= B)\; |phys> = 0,
\quad \mbox{and} \quad (ii)\; (p_x^2 + 2 m p_t) (\sim \dot B)\; |phys> = 0.
\end{array} \eqno(5.6)
$$
The above restrictions on the physical states are in agreement with
the Dirac's prescription for the consistent quantization of a physical
system endowed with the first-class constraints. In more precise words,
the operator form of the primary constraint $\Pi_E = B$ annihilates
the physical states of the theory
(i.e. $B\;|phys> = 0 \Rightarrow \Pi_E\; |phys> = 0$).
The requirement that this constraint
condition should remain intact w.r.t. ``time'' (i.e. $\tau$) evolution of
the system leads to the annihilation of the physical states by the
secondary-constraint (i.e. $(p_x^2 + 2 m p_t)\; |phys> = 0 \Rightarrow
\sim \dot B(= \dot \Pi_E)\; |phys> = 0$) of the theory. Thus, it is clear
that the physicality criteria $Q_{(a)b}\; |phys> = 0$ imply, in one stroke,
the annihilation of the physical states by both
the primary and the secondary constraints of the theory.

We dwell a bit on the derivation of the NC by exploiting (i) the
BRST transformations for the variables $x$ and $t$, and (ii) the
BRST cohomology connected with these transformations. To elaborate
it, let us re-express the off-shell nilpotent BRST transformations
for the space and time variables in (5.1) as $$
\begin{array}{lcl}
x(\tau) \rightarrow x^{(b)} (\tau) &=& x(\tau) + c (\tau)\; p_x
(\tau) \equiv  x(\tau) + s_{b} [x(\tau)], \nonumber\\ t(\tau)
\rightarrow t^{(b)} (\tau) &=& t(\tau) +  c (\tau)\; m \equiv t
(\tau)+ s_{b} [t(\tau)].
\end{array} \eqno(5.7)
$$ It is clear that (i) the transformed variables ($x^{(b)},
t^{(b)}$) and the original untransformed variables ($x, t$) belong
to the same cohomology class w.r.t. $s_b$ (because of the fact
that the BRST transformation $s_b$ is a nilpotent ($s_b^2 = 0$)
operator), and (ii) the space-time geometry is {\it commutative}
even in the BRST transformed frames (characterized by $x^{(b)}$
and $t^{(b)}$) because the Poisson brackets $\{x^{(b)}, t^{(b)}
\}_{(PB)} = \{ x, t \}_{(PB)} = 0$. Let us now focus on the
non-standard BRST-type transformations corresponding to (4.1),
namely;$$
\begin{array}{lcl}
x(\tau) \rightarrow \tilde x^{(b)} (\tau) &=&
 x (\tau) + c (\tau) \; m \equiv  x (\tau) + s_{b} [t(\tau)], \nonumber\\
t(\tau) \rightarrow \tilde t^{(b)} (\tau) &=& t (\tau) + c
(\tau)\; p_x (\tau) \equiv  t (\tau) + s_{b} [x(\tau)],
\end{array} \eqno(5.8)
$$ which can be thought of as the generalizations of the
non-standard symmetry transformations given in (4.1). It can be
readily seen that the above transformations (i.e. (5.8)) are also
cohomologically equivalent w.r.t. the nilpotent BRST
transformations $s_b$ {\it but} they lead to the existence of a
{\it noncommutative} geometry because of the NC in $\tilde
x^{(b)}$ and $\tilde t^{(b)}$ (i.e. $\{ \tilde x^{(b)}, \tilde
t^{(b)} \}_{(PB)} = c (\tau)$). The above statement is trivially
true (cf. (4.5)) for the case $p_x = m$ too (where the mass
parameter $m$ becomes noncommutative with $x(\tau)$). This
establishes ({\it vis-{\`a}-vis} transformations (5.7))
 the cohomological equivalence of the commutativity
and NC in the language of the nilpotent ($s_b^2 = 0$) BRST
transformations (5.1). The above arguments could be repeated with
the nilpotent anti-BRST transformations $s_{ab}$ as well. It will
be noted that, out of all the set of transformations in (5.8),
only a small subset of transformations corresponding to $p_x = m$,
turns out to be the symmetry transformation of the Lagrangian of
the present toy model as can be seen
 explicitly from (4.5).\\

\noindent {\bf 6 Quantum Groups on Phase Space and
Noncommutativity}\\

\noindent In this section, we very briefly discuss the connection
of our earlier work [19] on the dynamics of a $q$-deformed
particle, based on a consistently developed differential calculus
on the $q$-deformed phase space, and our present reparametrization
invariant NR toy model. To begin with, it can be noted that the
following relations on the phase space [19] $$\begin{array}{lcl}
x_i\; x_j = x_j \; x_i, \qquad p_i\; p_j = p_j\; p_i, \qquad x_i\;
p_j = q\; p_j\; x_i,
\end{array}\eqno(6.1)
$$ remain form invariant under the following transformations on
2N-dimensional phase space $$\begin{array}{lcl} x_i \to x^{(q)}_i
= A \;x_i + B \;p_i, \qquad p_i \to p^{(q)}_i = C\; x_i + D\; p_i,
\end{array}\eqno(6.2) $$
where (i) the original 2N-dimensional phase space (corresponding
to the N-dimensional configuration space characterized by $x_i$
(with $i = 1, 2, 3, .....N)$), is parametrized by $x_i$ and $p_i$,
(ii) the transformed phase space variables  ($x^{(q)}_i$ and
$p^{(q)}_i$) characterize the transformed co-tangent manifold,
(iii) the elements $A, B, C$ and $D$ belong to the $2 \times 2$
quantum group $GL_{q,p} (2)$ matrix that obey the following
braiding relations (see, e.g., [21]) $$
\begin{array}{lcl}
&& A B = p\; B A, \qquad C D = p\; D C, \qquad A C = q\; C A,
\qquad B D = q\; D B, \nonumber\\ && B C = (q/p)\; C B, \qquad A D
- D A = (p - q^{-1})\; B C = (q - p^{-1})\; C B,
\end{array}\eqno(6.3)
$$ where $q$ and $p$ are the non-zero complex numbers (i.e. $q, p
\in {\cal C}/\{0\}$) corresponding to the most general type of
quantum group deformation of the general linear group $GL (2)$ of
$2 \times 2$ non-singular matrices \footnote{It can be easily seen
that for $q = p = 1$, we obtain the commutative ordinary numbers
as elements $A, B, C$ and $D$ of an ordinary group $GL(2)$. This
is evident from the relations (6.3). Furthermore, for the
condition $p = q$, we retrieve the braiding relations for the
quantum group $GL_{q}(2)$ from (6.3).}, (iv) the exact form
invariance of (6.1) is assured only for the condition $pq = 1$ in
relations (6.3). In other words, the form invariance of
$q$-relations (6.1) is guaranteed under the quantum group
$GL_{q,q^{-1}} (2)$ which is a special case of the general quantum
group $GL_{q,p} (2)$ but is quite different from the single
parameter deformed quantum group $GL_{q}(2)$ (see, e.g., [21]),
(v) the relations in (6.3) reduce to the following relationships
for the quantum group $GL_{q,q^{-1}} (2)$, namely; $$
\begin{array}{lcl}
&& A B = q^{-1} B A, \qquad C D = q^{-1} D C, \qquad B C = q^2 C
B, \nonumber\\ && A C = q C A, \;\;\qquad \;\; B D = q D B,\;\;
\qquad \;\;A D = D A,
\end{array}\eqno(6.4) $$
which are found to be useful in the proof of the form invariance
of (6.1), (vi) in the above proof, the elements $A, B, C$ and $D$
are assumed to commute with the phase variables $(x_i, p_i)$ as
well as $(x^{(q)}_i, p^{(q)}_i)$,  and (vii) for our present toy
model, we have $x_i = (x_1, x_2) \equiv (x, t)$ and $p_i = (p_1,
p_2) \equiv (p_x, p_t)$. Thus, the phase space of our present
model is a four-dimensional co-tangent manifold that is
parametrized by the phase variables $x, p_x, t, p_t$.

To be precise, the transformations (6.2) correspond to two sets of
transformations for the two pairs of phase variables $(x, p_x)$
and $(t, p_t)$ that change under a couple of sets of the quantum
groups $GL_{q,q^{-1}} (2)$. These can be expressed explicitly in
the matrix form as $$
\begin{array}{lcl}
\left (
\begin{array}{c}
x\\ p_x\\
\end{array} \right )
\rightarrow \left (
\begin{array}{c}
x^{(q)}\\ p^{(q)}_x\\
\end{array} \right )
=
\left (
\begin{array}{cc}
A & B\\ C & D\\
\end{array} \right )
\left (
\begin{array}{c}
x\\ p_x\\
\end{array} \right ),
\end{array}\eqno(6.5)
$$
$$
\begin{array}{lcl}
\left (
\begin{array}{c}
t\\ p_t\\
\end{array} \right )
\rightarrow \left (
\begin{array}{c}
t^{(q)}\\ p^{(q)}_t\\
\end{array} \right )
=
\left (
\begin{array}{cc}
A & B\\ C & D\\
\end{array} \right )
\left (
\begin{array}{c}
t\\ p_t\\
\end{array} \right ),
\end{array}\eqno(6.6)
$$ where, as is obvious, the elements $A, B, C$ and $D$ obey the
$q$-algebraic relations (6.4). The canonical Poisson brackets
$\{x, p_x \}_{(PB)} = 1, \{t, p_t \}_{(PB)} = 1, \{x, t\}_{(PB)} =
0, \{p_x, p_t \}_{(PB)} = 0, \{x, p_t \}_{(PB)} = 0, \{t, p_x
\}_{(PB)} = 0$ etc., can be expressed concisely as $\{x_i, x_j
\}_{(PB)} = 0, \{p_i, p_j \}_{(PB)} = 0, \{x_i, p_j \}_{(PB)} =
\delta_{ij}$ where $i, j = 1, 2$ and $(x_1, x_2) = (x, t), (p_1,
p_2) = (p_x, p_t)$. These brackets have been consistently
generalized, with a specific choice of the $q$-deformed symplectic
structures on the phase space of the free NR particle, as  [19] $$
\begin{array}{lcl}
{\displaystyle \{ x_i, x_j \}^{(Q)}_{(PB)} = \{ p_i, p_j
\}^{(Q)}_{(PB)} = 0, \qquad \{ x_i, p_j \}^{(Q)}_{(PB)} =
\delta_{ij}, \qquad \{p_i, x_j \}^{(Q)}_{(PB)} = - q\;
\delta_{ij}}.
\end{array}\eqno(6.7)
$$ Using these $q$-canonical Poisson brackets and taking into
account the commutativity of the elements $A, B, C$ and $D$ with
the phase variables, we obtain the following NC $$
\begin{array}{lcl}
&& \{x^{(q)}_i, x^{(q)}_j \}^{(Q)}_{(PB)} = (A B - q B A)\;
\delta_{ij} \equiv (1 - q^2)\; A B \;\delta_{ij}, \nonumber\\ &&
\{p^{(q)}_i, p^{(q)}_j \}^{(Q)}_{PB} = (C D - q D C)\; \delta_{ij}
\equiv (1 - q^2)\; C D \;\delta_{ij}, \nonumber\\ && \{x^{(q)}_i,
p^{(q)}_j \}^{(Q)}_{PB} = ( A D - q B C )\; \delta_{ij} \equiv (D
A - q B C)\; \delta_{ij}.
\end{array}\eqno(6.8) $$ This shows that the NCs exist in
the spacetime as well as in the momentum coordinates. These NCs
lie in the transformed frames due to the quantum group
transformations (6.5) and (6.6) on the phase space. Thus, quantum
groups on the phase space do provide, in some sense, the most
general NCs on the phase space. These transformations, however,
become canonical transformations if we choose $q^2 = 1$ (i.e. $q =
\pm 1$) and the $q$-determinant of the $GL_{q,q^{-1}} (2)$ matrix
to be one. That is to say, the following choice$$
\begin{array}{lcl}
A D - q B C =  A D - q^3 C B = D A - q B C = D A - q^3 C B = 1,
\end{array}\eqno(6.9)
$$ entails upon the quantum group to reduce from $GL_{q,q^{-1}}
(2)$ to $SL_{q,q^{-1}} (2)$. It is easy to check that if we choose
$C = 0$ in (6.8), the NC in the momentum coordinates disappears
and only the NC persists in the spacetime coordinates. Similarly,
the other interesting possibilities can be explored by specific
choices of the elements of the quantum group.

Now let us focus on the gauge transformations in (3.3) which imply
that $x \to x^{(g)} = x + \xi p_x, t \to t^{(g)} = t + \xi m, p_x
\to p_x^{(g)} = p_x, p_t \to p_t^{(g)} = p_t$. It can be seen,
using the canonical brackets in the untransformed space (i.e.
$\{x_i, x_j\}_{(PB)} = \{p_i, p_j \}_{(PB)} = 0, \{ x_i, p_j
\}_{PB)} = \delta_{ij}$) that the following Poisson brackets are
valid in the transformed phase space $$
\begin{array}{lcl}
{\displaystyle \{ x^{(g)}_i, x^{(g)}_j \}_{(PB)} = 0, \quad
\{p^{(g)}_i, p^{(g)}_j \}_{(PB)} = 0, \quad \{x^{(g)}_i, p^{(g)}_j
\}_{(PB)} = \delta_{ij}},
\end{array}\eqno(6.10) $$ due to the gauge transformations written
in equation (3.3). This demonstrates that the gauge
transformations in (3.3) are the canonical transformations and
they correspond to the {\it commutative} geometry of the spacetime
structure. On the contrary, if we concentrate on the non-standard
symmetry transformations in (4.1), even for $p_x = m$, we have the
following brackets in the transformed frames $$
\begin{array}{lcl}
{\displaystyle \{ x^{(ns)}_i, x^{(ns)}_j \}_{(PB)} = \zeta
(\tau)\; \varepsilon_{ij}, \quad \{p^{(ns)}_i, p^{(ns)}_j
\}_{(PB)} = 0, \quad \{x^{(ns)}_i, p^{(ns)}_j \}_{(PB)} =
\delta_{ij}},
\end{array}\eqno(6.11)
$$ where we have used the basic canonical Poisson brackets of the
original untransformed phase space and we have exploited the NC of
the mass parameter with the space variable (i.e. $\{ x(\tau), m
\}_{(PB)} = 1$). Here totally antisymmetric Levi-Civita tensor in
two dimensions has been chosen such that $\varepsilon_{12} = + 1 =
\varepsilon^{12}$. A close look at (6.8) and (6.11) shows that,
under no possible restrictions, the NC of the transformations
(4.1) can be captured by the quantum group NC illustrated by the
Poisson bracket $\{x_i^{(q)}, x_j^{(q)} \}^{(Q)}_{(PB)} = (1 -
q^2)A B\;\delta_{ij}$ in (6.8). This is due to the fact that
$\delta_{ij}$ and $\varepsilon_{ij}$, present in (6.8) and (6.11),
respectively, have diametrically opposite mathematical properties
which cannot be reconciled in any manner.\\

\noindent {\bf 7 Conclusions}\\

\noindent In our present investigation, we have concentrated only
on a set of continuous symmetry transformations as a tool for the
discussion of the NC and commutativity in the context of the toy
model of a free massive NR particle. The reparametrization
symmetry, in some sense, is enforced on this model by treating the
``time'' parameter as a configuration space variable that depends
on a monotonically increasing evolution parameter $\tau$. One of
the key new features of our discussion is the result that the NC
for this toy model appears because of a set of non-standard
symmetry transformations (cf. (4.1),(4.2),(4.5)). In particular,
the symmetry transformations (4.5) can be understood in {\it two
different} ways which correspond to NC and commutativity. That is
to say (i) when (4.5) is derived as the special case of the gauge
transformations (3.3), it corresponds to a commutative geometry,
and (ii) when (4.5) is derived from the non-standard symmetry
transformations (4.1) and (4.2), it corresponds to the NC of
spacetime where (a) the mass parameter becomes noncommutative with
the space variable $x(\tau)$ alone, and (b) there exists a
restriction (i.e. $p_x = m$) in (4.2).

The reason behind the existence of the above kind of
``restricted'' symmetry transformations can be explained in the
language of the BRST cohomology. In fact, as it turns out, the
BRST transformations (cf. (5.8), (5.7)) that lead to the NC and
commutativity for this toy model, belong to the same cohomology
class w.r.t. the BRST charge $Q_b$ (or equivalently w.r.t. the
nilpotent BRST transformations $s_b$) \footnote{The BRST
cohomology, corresponding to the transformations (4.5), can be
readily discussed.}. However, there is a whole set of
transformations that is allowed by the BRST cohomology alone. The
transformations that correspond to the symmetry property of the
Lagrangian are restricted to be a subset of the above set of
cohomologically equivalent transformations in the phase space
where $p_x = m$. This explains, in a {\it new} way, the earlier
claims of [8,9] that the NC and commutativity owe their origin to
the gauge transformations for the general reparametrization
invariant theories that, of course, include our present toy model,
too (see, e.g.,[9]).

We have discussed the NC and commutativity of the spacetime
associated with the present toy model within the framework of the
quantum groups defined on the phase space of our present model. At
the level of the Poisson bracket structure, we have shown that the
commutativity (i.e. $\{ x^{(g)}_i, x^{(g)}_j \}_{(PB)} = 0$) of
the model, corresponding to the gauge symmetry transformations
(3.3), is equivalent to the transformations of the phase space
variables of the model under $SL_{q,q^{-1}} (2)$ where the
deformation parameter is restricted to be $q = \pm 1$. This kind
of restriction has also been found in [12,17,18]. We have been
able to establish theoretically, however, that the NC (i.e.
$\{x^{(ns)}_i, x^{(ns)}_j \}_{(PB)} = \zeta (\tau)
\varepsilon_{ij}$) associated with the non-standard symmetry
transformations in (4.1) cannot be captured by the NC of spacetime
structure associated with the quantum group (cf. (6.8)). This is
primarily due to the fact that: whereas the NC due to the quantum
group is associated with $\delta_{ij}$, the NC due to the
non-standard symmetry transformations in (4.1) is connected with
$\varepsilon_{ij}$. Both these mathematical quantities, having
their own specific properties, cannot be reconciled to each-other
simultaneously in any (mathematically consistent and correct)
manner.

Our present discussion has been generalized to the examples of the
reparametrization invariant (i) free massive relativistic particle
[22], and (ii) its interaction with the electromagnetic field in
the background [23]. It will be worthwhile to point out that the
models of the free (non-)relativistic particles and spinning
relativistic particles have also been considered in the framework
of quantum groups [16-19]. The latter framework is based on a
different kind of NC in spacetime structure. It will be an
interesting endeavour to find out some connections between the
above two approaches. The presence of the Snyder's NC has also
been shown in the context of the mechanical model of the two-time
physics [24]. It will be a nice venture to explore the possible
connection(s) of our approach with that of [24] for this very
interesting model in the language of continuous symmetry
properties. These are some of the future directions that are under
investigation at the moment and our results would be reported in
our forthcoming future publications [25].

\end{document}